\documentclass{article}

\usepackage{epsfig}
\title{Triaxial bifurcations of rapidly rotating spheroids}

\author{Ts.\ Dankova and  G.\ Rosensteel \\ 
Physics Department, Tulane University, \\
 New Orleans, Louisiana 70118 }

\begin{document}

\maketitle

\begin{abstract}
A rotating system, such as a star, liquid drop, or atomic nucleus, may rotate as an oblate spheroid about its symmetry axis or, if the angular velocity is greater than a critical value, as a triaxial ellipsoid about a principal axis. The oblate and triaxial equilibrium configurations minimize the total energy, a sum of the rotational kinetic energy plus the potential energy. For a star or galaxy the potential is the self-gravitating potential, for a liquid drop, the surface tension energy, and for a nucleus, the potential is the sum of the repulsive Coulomb energy plus the attractive surface energy. A simple, but accurate, Pad\'{e} approximation to the potential function is used for the energy minimization problem that permits closed analytic expressions to be derived. In particular, the critical deformation and angular velocity for bifurcation from MacLaurin spheroids to Jacobi ellipsoids is determined analytically in the approximation.
\end{abstract}

\section{INTRODUCTION}

A bifurcation is a qualitative shift in the character of the solutions to an equation. One of the best examples of a bifurcation is the transition from spheroidal to ellipsoidal shapes for rapidly rotating stars and galaxies. The equation for this problem is the energy minimization condition. Above a critical angular velocity, the total energy of a rigidly rotating triaxial ellipsoid is less than that of a spheroid.

The scientific investigations of the shape of self-gravitating systems form an important and 
fascinating segment of the history of physics \cite{TOD1873}. Although the best astronomical observations at the beginning of the eighteenth century indicated that the earth was prolate (Cassinian ovaloid), Sir Isaac Newton predicted that it was an oblate spheroid rotating about its symmetry axis. In 1738 Maupertuis organized a scientific expedition to 

Lapland whose aim was to resolve the controversy about the shape of the earth by making precise geodetic measurements.  He and his collaborator, Celsius, confirmed Newton's prediction and further substantiated the universal theory of gravitation. This success began the modern theory of the equilibrium shape of a rotating self-gravitating body.

Four years later, MacLaurin reported the rigorous relationship between the angular velocity $\omega$ and the eccentricity $e$ for the rigidly rotating spheroids that today bears his name,
\begin{equation}
\omega^{2}/ \pi G \rho = 2 (3 - 2 e^2 )\sqrt{1-e^2} \arcsin  (e)  /e^3 - 6 (1-e^2) / e^2
\end{equation}
where $G$ denotes the universal gravitational constant and $\rho$ is the constant density. An 
interesting feature of MacLaurin's formula is that the angular velocity attains a global maximum at an eccentricity $e \approx 0.93$.  For the small oblate eccentricities and slow rotations seen in the sun and planets, the shape of the bodies may be determined as a function of the angular velocity using elementary mechanics \cite{Boleman76}.

A century had elapsed before Jacobi in 1834 demonstrated that there were rigidly rotating triaxial equilibrium solutions in addition to the MacLaurin spheroids. The Jacobi ellipsoids bifurcate from the spheroids at $e \approx 0.81$. The angular velocity of a triaxial ellipsoid is less than the spheroidal maximum. At high angular velocities triaxial solutions are energetically preferred to oblate shapes, a counterintuitive result that illustrates dramatically the primacy of physical law to unreliable common sense.

There are other important results related to the shape of self-gravitating bodies. The discovery in 1924 of the instability of the bifurcation from ellipsoidal to pear shapes shattered the hope for a theory of binary star creation from embryonic ellipsoidal distributions of self-gravitating matter \cite{CAR24}. More recently precise measurement of the sun's oblate eccentricity using the solar disk sextant became important in part because its value distinguished Einstein's general theory of relativity from the competing Brans-Dicke theory of gravitation \cite{LYD96,DIC64}.

The objective of this article is to present a simple derivation of the MacLaurin to Jacobi bifurcation based on energy minimization. To attain the desired simplicity, the elliptic integral for the exact gravitational self-energy of a uniform density triaxial ellipsoid is approximated by a rational function of the ellipsoid's axes lengths. To emphasize the universality of this rotational dynamics problem, the oblate to triaxial bifurcation is determined for the similar problems of irrotational liquid drops and atomic nuclei \cite{COH74}. Here the surface area of an ellipsoid must be approximated by another Pad\'{e} function. Three science lessons are learned from the investigation: (1) What is a bifurcation and how can it be identified quantitatively?, (2) What is the physics of rotating isolated systems?, and (3) How is a complicated function like the self-energy approximated accurately by a rational Pad\'{e} function. The solutions to the exact bifurcation problems are found for self-gravitating systems in \cite{CHA69,LEB67} and for nuclei in \cite{COH74,ROS88}.

\section{Self-gravitating ellipsoids}

The first objective is to determine the total energy of an ellipsoid with semi-axes lengths 
$a_{1},a_{2},a_{3}$ that is rotating rigidly about a principal axis, say the first axis.  Since the system is isolated, the angular momentum $L$ is constant and aligned with the fixed rotation axis. The rotational kinetic energy is the square of the angular momentum divided by twice the moment of inertia ${\cal I}$ 
\begin{equation}
T = \frac{L^{2}}{2 {\cal I}} .
\end{equation}
For rigid body rotation, the moment of inertia of a uniform density ellipsoid of total mass M is 
${\cal I} = (M/5)(a_{2}^{2}+a_{3}^{2})$.

Size effects and scaling relations are clarified if dimensionless units are introduced. Moreover, the results may be applied then with equal ease to stars and galaxies that differ by many orders of magnitude in their linear dimensions, angular momenta, and periods. Consider an imaginary reference sphere of radius $R$ that has the same mass $M$ and volume $v$ as the 
ellipsoid, viz., $v = 4 \pi a_{1} a_{2} a_{3} / 3 = 4 \pi R^{3} / 3$. 
Define new dimensionless semi-axes lengths by $a=a_{1}/R, b=a_{2}/R, c=a_{3}/R$ for which the 
product $abc=1$. The  reference sphere's moment of inertia is ${\cal I}_{0} = 2 M R^{2} / 5$, 
and the ellipsoid's is ${\cal I} = {\cal I}_{0} (b^{2}+c^{2}) / 2$.

The gravitational self-energy of a mass distribution with density $\rho$ is the total potential energy of gravitational attraction. It is given by a double integral over all pairs of infinitesimal masses located at $\vec{r}$ and $\vec{r}^{\, \prime}$ 
\begin{equation}
V = - \frac{G}{2} \int \int \frac{\rho(\vec{r}) d^{3}r \, \rho(\vec{r}^{\, \prime}) d^{3}r^{\prime}}{|\vec{r}-\vec{r}^{\, \prime}|} . \label{exactself}  
\end{equation}
Division by two is required because of the integrand's double counting of the infinitesimal mass pairs. 

The gravitational self-energy of the uniform reference sphere is evaluated to be
$V = -V_{0} = -(3/5) G M^{2} / R$. With somewhat more difficulty, MacLaurin evaluated the double integral for an oblate spheroid. If $b=c>a$, the gravitational self-energy is 
\begin{equation}
V = -V_{0} (1-e^{2})^{1/6}\frac{\arcsin(e)}{e} 
= \left\{ \begin{array}{ll} -V_{0} [1 -\frac{1}{45}e^{4} -\frac{62}{2835}e^{6} -\ldots ] 
& \mbox{\rm for $e << 1$} \\
-V_{0} \frac{\pi}{2}(1-e^{2})^{1/6} & \mbox{for $e \rightarrow 1$} \end{array} \label{MacLaurin}
\right. ,
\end{equation} 
where the eccentricity $e=\sqrt{1-a^{2}/c^{2}}$.

The self-energy for a triaxial ellipsoid cannot be evaluated in terms of elementary functions, but the double integral may be reduced to an elliptic integral\cite{CHA69}
\begin{equation}
V = -(V_{0}/2) \int_{0}^{\infty }\frac{\,du}{\Delta}
\left[ \frac{a^{2}}{a^2+u} +  \frac{b^{2}}{b^2+u}+  
\frac{c^{2}}{c^2+u} \right] , \label{ellipticintegral}
\end{equation}
where $\Delta = \sqrt{ (a^2+u) (b^2+u) (c^2+u) }$.

Instead of working directly with this elliptic integral, it is both technically simpler and more physically illuminating to approximate it. Although the infinite Taylor series expansion about the sphere $a=b=c=1$ is identical to the original analytic function, a truncation to a polynomial introduces serious errors for large axes lengths. The reason is that the polynomial violates an important qualitative feature of the exact function. Observe that the ellipsoid's exact gravitational self-energy, Eq.(\ref{exactself}) or Eq.(\ref{ellipticintegral}), is a function of the axes lengths satisfying three general properties: (1) it is a symmetric function of the axes lengths, (2) it is homogeneous of degree $-1$ in the axes lengths, and (3) it equals the reference sphere value when the ellipsoid's axes lengths are equal. A superior approximation is achieved if all three properties are respected. Although the monomial terms of a truncated Taylor series violate the homogeneity property, an elementary function satisfying all three properties is
\begin{equation}
V_{1} =  -V_{0} \frac{3}{a+b+c} .
\end{equation}
Although $V_{1}$ is a good approximation to the exact self-energy of an ellipsoid, an even better approximation is the two term expression
\begin{equation}
V_{2} = -V_{0} \left[ \frac{4}{5}\frac{3}{a+b+c} + \frac{1}{5}
\frac{a+b+c}{ab+bc+ca} \right].
\end{equation}
The weights for the two terms, $4/5$ and $1/5$, are chosen to attain agreement with the exact MacLaurin oblate spheroid energy, Eq. (\ref{MacLaurin}), up to fourth order  in the eccentricity,
\begin{equation}
V_{2} = \left\{ \begin{array}{ll} -V_{0} [1 - \frac{1}{45}e^{4} -\frac{7}{324}e^{6} - \ldots ]
& \mbox{\rm for $e << 1$} \\
-V_{0} \frac{8}{5}(1-e^{2})^{1/6} & \mbox{for $e \rightarrow 1$} \end{array} . \right.
\end{equation}

A rational approximation, i.e., a ratio of polynomials such as $V_{1}$ and $V_{2}$, to a function is called a Pad\'{e} approximation \cite{STO80}. Note the close agreement between the exact and approximate expressions even beyond the fitted fourth order in the eccentricity. The coefficients of $e^{6}$ in the exact and approximate series differ only by about $1\%$,  $62/2835 \approx 0.0219$ versus $7/324 \approx 0.0216$. In the limit of an oblate spheroid (flat pancake), $e\rightarrow 1$, both the exact and approximate formulae tend to zero as $(1-e^{2})^{1/6}$, and their coefficients differ only by about $2$\%, $\pi/2 \approx 1.57$ compared to $8/5 =1.60$. The algebraic manipulations and Taylor expansions were done here and elsewhere in this paper with the assistance of a computer algebra program.

Because of centrifugal stretching, the rotation axis $a$ is shorter than $b$ and $c$, and the relevant domain to describe rotational motion around the $a$-axis is $b^2 c >1$ and $b c^2 >1$. In Figure 1 a contour plot of the percentage error for the Pad\'{e} approximation $V_{2}$ is shown in the $b-c$ plane for $0.5 < b, c < 2.0$. This domain encompasses the shapes of self-gravitating systems found in nature. The error there is very small -- less than one percent. In this same rectangular domain, the percentage error in the derivatives of the potential energy, i.e., the Chandrasekhar tensor, is also less than one percent.

The errors are largest for prolate spheroids that lie outside the physical domain, cf.\ Eq.(\ref{Macangmom}) below. In Figure 2 the potential energy and its 
Pad\'{e} approximant are plotted versus the eccentricity of the prolate spheroid. The differences between the exact and approximate energies cannot be seen on the scale $0 < e < 1$ except near the infinitely thin elongated prolate spheroid $e\rightarrow 1$. The difference between the exact potential and its approximation can be seen in the magnified interval $0.9990 < e < 1$. The exact potential has a logarithmic singularity as $e \rightarrow 1$ and the approximation diverges from it. This divergence is far outside the physically relevant domain. Were the study of extreme prolate shapes important, an approximation with the correct logarithmic divergence should be adopted, e.g., 
\begin{equation}
V_{\ln} = \frac{3}{a+b+c} \left[ 1 + \frac{1}{5} \ln\left( \frac{(a+b+c)^2}{3(ab+bc+ca)} \right) \right] . 
\end{equation}

We would like to point out that there is an interesting phenomenon in quantum mechanics that contradicts the classical expectation of centrifugal stretching and oblate-like rotating shapes. At finite temperature, certain nuclei can rotate as prolate spheroids about their symmetry axes, $a > b=c$ \cite{Goodman}.

Let the energy and angular momentum be given in dimensionless form by $\epsilon = E/V_{0}$ and $\lambda  =L/\sqrt{I_{0}V_{0}}$, respectively.  Thus the total energy of a uniform self-gravitating ellipsoid is approximated accurately by the dimensionless expression
\begin{equation}
\epsilon (b,c) = \frac{{\lambda}^2}{b^2+c^2} -  \frac{4}{5}\frac{3}{a+b+c} - \frac{1}{5}
\frac{a+b+c}{ab+bc+ca} , \mbox{\ where } a=\frac{1}{bc} .
\end{equation}

For a uniform self-gravitating ellipsoidal mass rotating rigidly with angular momentum $\lambda$, the equilibrium shape is given by energy minimization
\begin{equation}
0 = \frac{\partial \epsilon}{\partial b} = \frac{\partial \epsilon}{\partial c} .
\end{equation}
These two equations are equivalent to
\begin{eqnarray}
0 & =  & \frac{\partial \epsilon}{\partial b} - \frac{\partial \epsilon}{\partial c} \nonumber \\
& = & (c-b) \left[\frac {2\,\lambda^{2}}{\left(c^{2}+b^{2}\right)
 ^{2}}+\frac {\left(b^{2}\,c^{3}+b^{3}\,c^{2}-c^{2}
 -b^{2}\right)}{5\,\left(b^{2}\,c^{2}+c+b\right)^{2}} \right. \nonumber \\ 
& & \hspace{1.5cm} \left. -\frac {12}{5\,\left(b\,c^{2}+b^{2}\,c+1\right)^{2}} \right] \label{diff} \\
0 & = & \frac{\partial \epsilon}{\partial b} + \frac{\partial \epsilon}{\partial c} \nonumber \\
& = &  -\frac {2\,\left(c+b\right)\,\lambda^{2}}{\left(c^{2}+b^{2}\right)
 ^{2}}+\frac {b^{2}\,c^{4}-c^{3}+b^{4}\,c^{2}-b\,c^{2}-b^{2}\,c-b^{3}
 +2}{5\,\left(b^{2}\,c^{2}+c+b\right)^{2}} \nonumber \\
& & \hspace{1.5cm}+\frac {12\,\left(2\,b^{2}
 \,c^{2}-c-b\right)}{5\,\left(b\,c^{2}+b^{2}\,c+1\right)^{2}} \label{sum}
\end{eqnarray}

One class of simultaneous solutions to the equilibrium equations is the MacLaurin spheroids for which $c=b$. In this case, Eq.\,(\ref{diff}) is trivially satisfied and Eq.\,(\ref{sum}) determines the angular momentum as a function of the long axis
\begin{equation}
 \lambda^{2}=\frac {2\,c\,\left(c^{3}-1\right)\,\left(16\,c^{9}+48\,c^{6}+45\,c^{3}-1\right)}
{5\,\left(c^{3}+2 \right)^{2}\,\left(2\,c^{3}+1\right)^{2}} . \label{Macangmom}
\end{equation}
Note that prolate solutions $(c<1)$ are excluded since the angular momentum cannot be imaginary. Although the angular momentum monotonically increases as a function of $c$, the angular velocity, $\sqrt{I_{0}/V_{0}}\, \omega = 2\lambda/(b^2+c^2) = \lambda/c^2$, attains a maximum at $c\approx 1.393$ or an eccentricity $e\approx 0.929$, a one-tenth of one percent error from the exact value  $e=0.930$.
The energy of a MacLaurin spheroid is
\begin{equation}
\epsilon =  -\frac {16\,c^{12}+104\,c^{9}+177\,c^{6}+107\,c^{3}+1}{5\,c\,
 \left(c^{3}+2\right)^{2}\,\left(2\,c^{3}+1\right)^{2}}
\end{equation}

A second class of equilibrium solutions is the Jacobi ellipsoids for which $b \neq c$. If Eq.\,(\ref{diff}) is divided by $c-b$, Eq.({\ref{sum}) divided by $c+b$, and the two new equations added together, then the angular momentum is eliminated and the resulting condition for the shape of an equilibrium triaxial ellipsoid is 
\begin{eqnarray}
0 & = & a^{4}\,\xi^{5} + a^{2}\left(2\,a^{3}-1\right)\,\xi^{4}+
a^{3}\left(13\,a^{3}-4\right)\,\xi^{3}+ a\left(6\,a^{3}+1\right)\,\xi^{2} \nonumber \\
& & -a^{2}\left(4\,a^{3}+10\right)\,\xi-a^{6}+a^{3}-12 , \label{shape}
\end{eqnarray}
where $\xi = c+b$ and $a=1/(bc)$.

The bifurcation of MacLaurin spheroids to Jacobi ellipsoids is determined by the solutions to the Jacobi shape equation in the limit $b\rightarrow c$
\begin{equation}
0 = 24\,c^{12}+20\,c^{9}-57\,c^{6}-96\,c^{3}+1 . 
\end{equation}
Hence the bifurcation is at $b=c\approx 1.210$, $a\approx 0.683$, or an eccentricity $e=0.826$. The exact bifurcation point, first determined by Mayer in 1842, is $e=0.813$; the error due to the Pad\'{e} approximation to the potential energy is about $1.5\%$. In Figure 3 the solutions to the Jacobi shape equation (\ref{shape}) are plotted in the $b-c$ plane. The intersection of the diagonal line $b=c$, corresponding to MacLaurin spheroids, with the Jacobi triaxial ellipsoidal curve is the bifurcation point. In Figure 4 the square of the angular velocity $\omega^2$, in units of $V_{0}/I_{0}$, is plotted versus the eccentricity $e$. The triaxial ellipsoids rotate more slowly than the spheroids beyond the bifurcation point.

\section{Liquid Drops}
Consider next a water or oil droplet with an ellipsoidal boundary that is rotating about a principal axis, say the $x$-axis. Microgravity experiments on the space shuttle are an ideal setting for measuring the properties of rotating fluid droplets\cite{WANG96}. For inviscid  incompressible irrotational flow, the ellipsoidal droplet's moment of inertia equals ${\cal I} = I_{0} (b^2-c^2)^2/(2(b^2+c^2))$ instead of the rigid body value.

The potential energy is the surface energy, defined as the product of the surface tension $\gamma$ times the area of the ellipsoid. For the reference sphere the surface energy is $V_{0}^{S} = 4\pi R^2 \gamma$. The surface area of an oblate spheroid is 
\begin{eqnarray}
V & = & \frac{1}{2} V_{0}^{S} (1-e^2)^{2/3} \left[ \frac{1}{1-e^2} + \frac{\tanh^{-1} e}{e}  \right]  \\
& = &  \left\{ \begin{array}{ll} V_{0}^{S} [1 + \frac{2}{45}e^4 + \frac{136}{2835}e^6 + \ldots ] 
& \mbox{\rm for $e << 1$} \\
V_{0}^{S} \frac{1}{2} (1-e^2)^{-1/3} & \mbox{for $e \rightarrow 1$} \end{array} \nonumber \label{oblatesurface}
\right.
\end{eqnarray}

The exact surface energy of a triaxial ellipsoid cannot be expressed in terms of elementary functions, but rather requires an elliptic integral \cite{ROS67}
\begin{equation}
V = \frac{V_{0}^{S}}{2} \int_{0}^{\infty }\frac{\,dt}{\chi}
\left[ \frac{1}{a^2+t^2} +  \frac{1}{b^2+t^2}+  
\frac{1}{c^2+t^2} \right]
\end{equation}
where $\chi = \sqrt{(a^2+t^2) (b^2+t^2) (c^2+t^2)}$.

To approximate this elliptical integral for the surface area of a triaxial ellipsoid, a strategy similar to the gravitational self-energy is successful. An ellipsoid's surface energy is a function of the axes lengths satisfying three properties: (1) it is a symmetric function of the axes lengths, (2) it is homogeneous of degree $+2$ in the axes lengths, and (3) it equals the reference sphere value when the ellipsoid's axes lengths are equal. A rational function with these three properties is the two term approximation
\begin{equation}
V_{2}^{S} =  V_{0}^{S} \left[ \frac{8}{15}(ab+bc+ca) - \frac{3}{5}(abc)^{2/3} \right] .
\end{equation}
The weights for the two terms, $8/15$ and $-3/5$, are chosen to attain agreement with the exact surface area of an oblate spheroid to fourth order in the eccentricity,
\begin{equation}
V_{2}^{S} = \left\{ \begin{array}{ll} V_{0}^{S} [1 + \frac{2}{45}e^{4} +\frac{19}{405}e^{6} - \ldots ]
& \mbox{\rm for $e << 1$} \\
V_{0}^{S} \frac{8}{15}(1-e^{2})^{-1/3} & \mbox{for $e \rightarrow 1$} \end{array} . \right.
\end{equation}

The exact and approximate expressions are close even beyond the fitted fourth order. The coefficients of $e^{6}$ in the exact and approximate series differ only by about $2\%$,  $136/2835 \approx 0.0480$ versus $19/405 \approx 0.0469$. In the limit, $e\rightarrow 1$ (cr\^{e}pe), both the exact and approximate formulae approach infinity as $(1-e^{2})^{-1/3}$, with coefficients differing by $7$\%, $0.50$ compared to $8/15 =0.53$. The singularity cannot be incorporated into a finite degree Taylor polynomial. To even qualitatively reproduce this property of the surface area, a Pad\'{e} approximation is necessary.

The total energy $\epsilon$, in units of the reference surface energy $V_{0}^{S}$, is 
\begin{equation}
\epsilon (b,c) = \frac{{(b^2+c^2) \lambda}^2}{(b^2-c^2)^2} +  \frac{8}{15}(ab+bc+ca) - \frac{3}{5} , \mbox{\ where } a=\frac{1}{bc} 
\end{equation} 
and the angular momentum $\lambda$ is in units of $\sqrt{I_{0}V_{0}^{S}}$.

Energy minimization yields two equations, one for the shape of the equilibrium fluid drops,
\begin{equation}
0 = a^{4}\,\xi^{5}-a^{2}\,\xi^{4}-5\,a^{3}\,\xi^{3}+4\,a\,\xi^{2}+8\,a^{2}\,\xi- 8, \label{dropshape}
\end{equation}
and the other for the angular momentum
\begin{equation}
\lambda^{2}=\frac{4}{15a^2} \left( -3a^3\xi^3+4a\xi^2+8a^2\xi-8 \right) ,
\end{equation}
where $\xi = b+c$, $a=1/bc$. 

For an irrotational drop no spheroidal solution exists save for the sphere itself at zero angular momentum. As soon as the drop begins to rotate, it immediately goes triaxial. In Figure 5 the equilibrium shapes, the solutions to Eq.\,(\ref{dropshape}), are plotted in the $b-c$ plane.

\section{Nuclear liquid drop model}
In a seminal paper in 1939, N.\ Bohr and J.A.\ Wheeler noted that an atomic nucleus has properties in common with a charged rotating incompressible liquid droplet whose stability is determined by a competition between the attractive surface energy and the repulsive effects of the Coulomb force and centrifugal stretching \cite{BOH39}. The surface energy, which approximates the effect of the short range nuclear interaction, is the same as a water droplet's energy except that the surface tension takes a different numerical value. The $Z$ protons produce the long range Coulomb potential. The repulsive Coulomb self-energy is similar to the attractive self-gravitating energy except for a sign change and different reference energy $V_{0}^{C} = (3/5)k_c Q^{2}/R$ where $k_c$ denotes the Coulomb constant, $Q=Ze$ is the nuclear charge, and $R$ is the radius of the reference sphere. The nuclear radius is proportional to the cube root of the mass number $A$, the total number of protons and neutrons, $R=r_0 A^{1/3}$, $r_0 = 1.2 F$. A microscopic justification, based on the Yukawa pion exchange potential, for the surface energy approximation to the strong force in nuclei is given by Gauthier and Sherrit \cite{GAU91}. Moreover, the analogy between liquid drops and atomic nuclei is supported by collision experiments \cite{MEN86}.

A basic conclusion of the Bohr-Wheeler fission model is that a nonrotating nucleus becomes unstable if the fissility $x=V_{0}^{C}/2V_{0}^{S}$ is greater than one. To respect this bound increasing numbers of neutrons must be supplied to the nucleus. But these extra neutrons are not stable against $\beta$ decay. Hence the periodic table terminates around $Z\sim 100$. For a  nonrotating spherical nucleus the energy simplifies to $V(R) = V_{0}^{S}+V_{0}^{C}$ and equilibrium is attained when $0 = R\, V^{\prime}(R) = 2 V_{0}^{S} - V_{0}^{C}$ or, equivalently, the fissility $x=1$. Using this article's approximations for the potentials of deformed ellipsoids, the stability will be investigated now for rotating ellipsoidal nuclei.

In units of the reference surface energy, the total energy of a rigidly rotating atomic nucleus is modeled by
\begin{eqnarray}
\epsilon (b,c) & = & \frac{{\lambda}^2}{b^2+c^2} + 2\, x\, \left[ \frac{4}{5}\frac{3}{a+b+c} + \frac{1}{5}
\frac{a+b+c}{ab+bc+ca} \right] \nonumber \\
& & + \frac{8}{15}(ab+bc+ca) - \frac{3}{5} , \mbox{\ where } a=\frac{1}{bc} .
\end{eqnarray}

There are both oblate spheroidal and triaxial ellipsoidal solutions to the two energy minimization equations, $\partial \epsilon / \partial b = \partial \epsilon / \partial c = 0$. For the oblate spheroids $(b=c)$, the angular momentum is 
\begin{equation}
\lambda^{2} = \frac{16}{15}c\,\left(c^{3}-1\right) \left[ 1 - \frac{3}{4} x 
\frac{16 c^{9} + 48 c^{6} + 45 c^{3} -1}{(1+2 c^{3})^{2}(2+c^{3})^{2}} \right] . 
\end{equation} 

If $c\geq 1$ and $0<x<1$, the square of the angular momentum is nonnegative. A key prediction is an upper bound on the spheroid's angular velocity
\begin{equation}
\lim_{c\rightarrow\infty} \frac{{\cal I}_{0}}{V_{0}^{S}}\, \omega^2 = \frac{16}{15} .
\end{equation}

The triaxial solutions obey the equilibrium shape condition
\begin{eqnarray}
0 & =  & 4\,\left(\xi+a\right)^{2}\,\left(a^{2}\,\xi+1\right)^{2}\,
 \left(a^{2}\,\xi^{2}-\xi-a\right) \nonumber \\
& & - 3x\,\left[ a^{4}\,\xi^{5} + a^{2}\left(2\,a^{3}-1\right)\,\xi^{4}+
a^{3}\left(13\,a^{3}-4\right)\,\xi^{3}\right. \\
& & \left. + a\left(6\,a^{3}+1\right)\,\xi^{2} -a^{2}\left(4\,a^{3}+10\right)\,\xi-a^{6}+a^{3}-12\right] \nonumber
\end{eqnarray}
where $\xi=c+b$ and $a=1/bc$. The bifurcation from oblate spheroids to triaxial ellipsoids is defined by this shape equation in the limit $b\rightarrow c$
\begin{equation}
x=\frac {4\,\left(c^{3}+2\right)^{2}\,\left(2\,c^{3}-3\right)\,\left(2\,c^{3}+1\right)^{2}}
{3\,\left(24\,c^{12}+20\,c^{9}- 57\,c^{6}-96\,c^{3}+1\right)}
\end{equation}
For spheroids $b=c=1$, the shape equation implies $x=1$, as expected. In Figure 6 the equilibrium shapes are drawn in the $b-c$ plane for three values of the fissility.

\section{CONCLUSION}
A simple approximate solution to the Jacobi bifurcation problem for a uniform self-gravitating rotating fluid with an ellipsoidal boundary is achieved in this article by using a Pad\'{e} approximation to the exact potential energy. To enhance agreement with the exact problem, terms of higher degree in the rational approximation are required with a concomitant loss of simplicity. However the exact problem is a mathematical problem, not a physical one. Real stars and galaxies are not uniform density fluids. The error in the Pad\'{e} approximation to the exact self-gravitating energy is less than the error made in imposing a constant density approximation on a real star or galaxy. Using the exact elliptic integral instead of the Pad\'{e} function is poor modeling of the physical world. Indeed, part of the art of modeling is recognizing inherent model limitations and adapting the mathematical complexity to resonate harmoniously with the physical assumptions. Failure to do so may even obscure the model's domain of application to nature. Similar remarks apply to the nuclear liquid drop model in which the surface energy approximates the short range hadronic interaction.

\vspace{2cm}

\noindent {\it Acknowledgment}. We thank J.D. Garrett and A.L. Goodman for their helpful remarks.

\newpage
\begin{figure}[htbp]\centering
\mbox{\psfig{figure=fig1.eps,height=5in,width=5in,clip=}}
\caption{Contour plot of the percentage error in the Pad\'{e} approximation $V_{2}$ to the exact self-gravitating potential energy.}
\end{figure}

\newpage
\begin{figure}[htbp]\centering
\mbox{\psfig{figure=fig2.eps,height=5in,width=5in,clip=}}
\caption{Exact and approximate potentials of self-gravitating prolate spheroids are plotted versus the eccentricity. The interval near $e=1$ is magnified to show the logarithmic divergence of the exact potential from the rational approximation.}
\end{figure}

\newpage
\begin{figure}[htbp]\centering
\mbox{\psfig{figure=fig3.eps,height=5in,width=5in,clip=}}
\caption{The equilibrium self-gravitating ellipsoids are drawn in the $b-c$ plane. The triaxial Jacobi ellipsoid curve bifurcates from the Maclaurin spheroid line at $b=c\approx 1.210$.}
\end{figure}

\newpage
\begin{figure}[htbp]\centering
\mbox{\psfig{figure=fig4.eps,height=5in,width=5in,clip=}}
\caption{The square of the angular velocity is plotted versus the eccentricity. The spheroids bifurcate to Jacobi ellipsoids rotating more slowly.}
\end{figure}

\newpage
\begin{figure}[htbp]\centering
\mbox{\psfig{figure=fig5.eps,height=5in,width=5in,clip=}}
\caption{The equilibrium triaxial liquid drops bifurcate from the sphere $b=c=1$.}
\end{figure}

\newpage
\begin{figure}[htbp]\centering
\mbox{\psfig{figure=fig6.eps,height=5in,width=5in,clip=}}
\caption{The bifurcation from oblate spheroids to triaxial ellipsoids in atomic nuclei depends on the fissility $x$. If $x=1$ the bifurcation occurs at the sphere $b=c=1$.}
\end{figure}

\end{document}